\documentclass[twocolumn,preprintnumbers,showpacs,amsmath,amssymb]{revtex4}
\usepackage{graphicx}
\usepackage{amssymb}

\begin{document}
\title{Simplified quantum bit commitment using single photon nonlocality}
\author{Guang Ping He}
\email{hegp@mail.sysu.edu.cn}
\affiliation{School of Physics and Engineering, Sun Yat-sen University, Guangzhou 510275, China}

\begin{abstract}
We simplified our previously proposed quantum bit commitment (QBC) protocol
based on the Mach-Zehnder interferometer, by replacing symmetric beam
splitters with asymmetric ones. It eliminates the need for random sending
time of the photons; thus, the feasibility and efficiency are both improved.
The protocol is immune to the cheating strategy in the Mayers-Lo-Chau no-go
theorem of unconditionally secure QBC, because the density matrices of the
committed states do not satisfy a crucial condition on which the no-go
theorem holds.
\end{abstract}



\maketitle


\section{Introduction}

Quantum bit commitment (QBC) \cite{qi365,qi43} is a two-party cryptography
including two phases. In the commit phase, Alice (the sender of the
commitment) decides the value of the bit $b$ ($b=0$ or $1$) that she wants
to commit, and sends Bob (the receiver of the commitment) a piece of
evidence, e.g., some quantum states. Later, in the unveil phase, Alice
announces the value of $b$, and Bob checks it with the evidence. An
unconditionally secure QBC protocol needs to be both binding (i.e., Alice
cannot change the value of $b$ after the commit phase) and concealing (Bob
cannot know $b$ before the unveil phase) without relying on any
computational assumption.

QBC is recognized as an essential primitive for quantum cryptography, as it
is the building block for quantum multi-party secure computations and more
complicated \textquotedblleft post-cold-war era\textquotedblright\
multi-party cryptographic protocols \cite{qi75,qi139}. Unfortunately, it is
widely accepted that unconditionally secure QBC is impossible \cite{qi74}-%
\cite{qbc32}, despite of some attempts toward secure ones (e.g., \cite%
{HeJPA,HePRA,Hearxiv,Hearxiv2,qbc75,qi92} and the references therein). This
result, known as the Mayers-Lo-Chau (MLC) no-go theorem, was considered as
putting a serious drawback on quantum cryptography. (Note that
cheat-sensitive QBC \cite{qi150,qbc52,qbc89} is not covered, as its security
goal is defined differently from that of the QBC studied in the current
paper.)

Very recently, we proposed a QBC protocol using orthogonal states \cite%
{HeJPA}, where the density matrices do not satisfy a crucial condition on
which the MLC no-go theorem holds. 
The protocol is based on the Goldenberg-Vaidman (GV) quantum key
distribution (QKD) scheme \cite{qi858}, which makes use of the Mach-Zehnder
interferometer involving symmetric beam splitters. Koashi and Imoto \cite%
{qi917} pointed out that the GV QKD scheme can be simplified by replacing
the symmetric beam splitters with asymmetric ones. Here we will apply the
same idea to propose a simplified version of our QBC protocol, so that it
can be more feasible and efficient.

In the next section, we introduce some basic notations and settings used
throughout the paper. The Koashi-Imoto (KI) QKD scheme will be briefly
reviewed in section III. Then we propose our QBC protocol in section IV, and
analyze its security in section V. The feasibility will be discussed in
section VI. Section VII summarizes the conclusion and gives some remarks. An
example of the technical attack mentioned in section VI will be provided in
the appendix.

\section{Notations and settings}

Generally, in both QKD and QBC the two participants are called Alice and
Bob. But similar to \cite{HeJPA}, in our current QBC protocol, the behavior
of Bob is more like that of the eavesdropper rather than the Bob in QKD. To
avoid confusion, here we use the names in the following way. In QKD, the
sender of the secret information is called Alice, the receiver is renamed as
Charlie instead of Bob, and the external eavesdropper is called Eve. In QBC,
the sender of the commitment is Alice, the receiver is Bob, and there is no
Eve since QBC merely deals with the cheating from internal dishonest
participants, instead of external eavesdropping.

As our main interest is focused on the theoretical possibility of secure
QBC, we will only consider the ideal case where no transmission error occurs
in the communication channels, nor there are detection loss or dark counts,
etc.

\section{The Koashi-Imoto QKD scheme}


\begin{figure*}[tbp]
\includegraphics{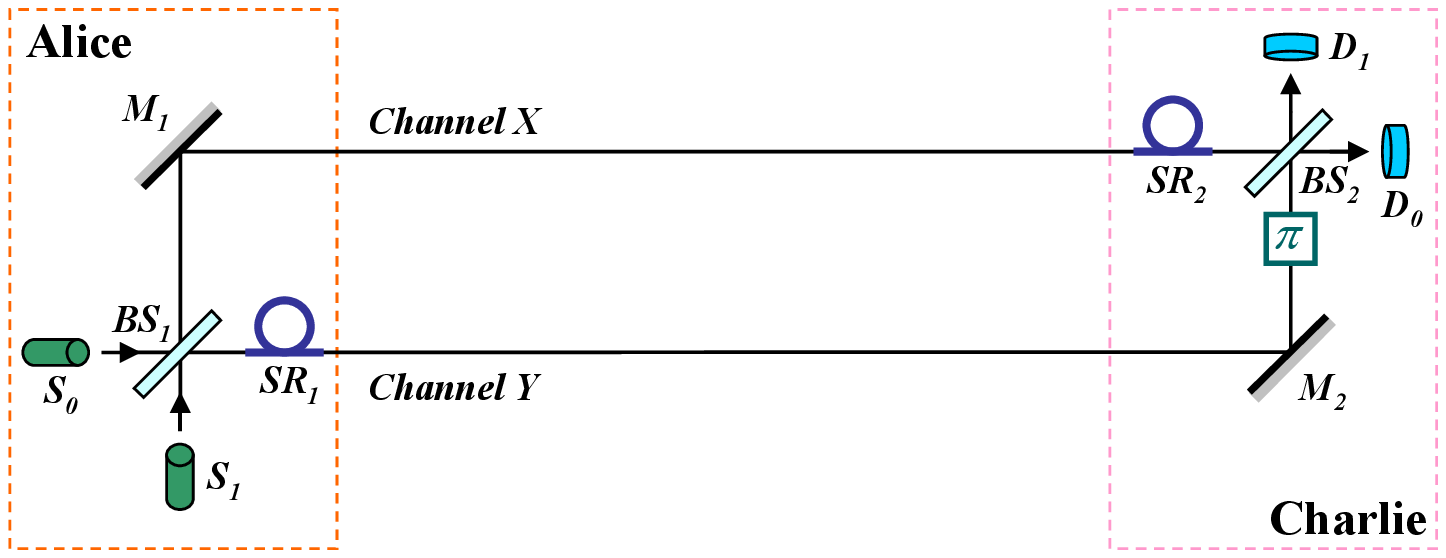}
\caption{Diagram of the experimental implementation of the Koashi-Imoto QKD
scheme \protect\cite{qi917}. The state of a photon produced by the source $%
S_{0}$ ($S_{1}$) will become $\left\vert \Psi _{0}\right\rangle =\protect%
\sqrt{T}\left\vert 0\right\rangle _{X}\left\vert 1\right\rangle _{Y}-i%
\protect\sqrt{R}\left\vert 1\right\rangle _{X}\left\vert 0\right\rangle _{Y}$%
\ ($\left\vert \Psi _{1}\right\rangle =\protect\sqrt{T}\left\vert
1\right\rangle _{X}\left\vert 0\right\rangle _{Y}-i\protect\sqrt{R}%
\left\vert 0\right\rangle _{X}\left\vert 1\right\rangle _{Y}$) after passing
the asymmetric beam splitter $BS_{1}$ with reflectivity $R$ and
transmissivity $T$. The two wave packets of the same photon are sent through
channels $X$ and $Y$ respectively. When no eavesdropper presented, the
storage rings $SR_{1}$, $SR_{2}$, the mirrors $M_{1}$, $M_{2}$ and the phase
shifter $\protect\pi$ will ensure the complete apparatus work as a
Mach-Zehnder interferometer, so that $\left\vert \Psi _{0}\right\rangle $\ ($%
\left\vert \Psi _{1}\right\rangle $) will be detected by the detector $D_{0}$
($D_{1}$) with certainty.}
\label{fig:epsart}
\end{figure*}


Our QBC proposal is inspired by the KI QKD scheme \cite{qi917}, which makes
use of the Mach-Zehnder interferometer illustrated in\ FIG. 1. Let $R$ and $T
$ denote the reflectivity and transmissivity of the asymmetric beam
splitters $BS_{1}$\ and $BS_{2}$, with $R+T=1$\ and $R\neq T$. Alice encodes
the bit values $0$ and $1$ she wants to transmit to Charlie, respectively,
using two orthogonal states%
\begin{eqnarray}
0 &\rightarrow &\left\vert \Psi _{0}\right\rangle \equiv \sqrt{T}\left\vert
0\right\rangle _{X}\left\vert 1\right\rangle _{Y}-i\sqrt{R}\left\vert
1\right\rangle _{X}\left\vert 0\right\rangle _{Y},  \nonumber \\
1 &\rightarrow &\left\vert \Psi _{1}\right\rangle \equiv \sqrt{T}\left\vert
1\right\rangle _{X}\left\vert 0\right\rangle _{Y}-i\sqrt{R}\left\vert
0\right\rangle _{X}\left\vert 1\right\rangle _{Y}.  \label{eqpsi}
\end{eqnarray}%
Here $\left\vert m\right\rangle _{j}$ is the $m$ photon Fock state for the
arm $j=X,Y$. That is, each $\left\vert \Psi _{0}\right\rangle $\ or $%
\left\vert \Psi _{1}\right\rangle $\ is split into two localized wave
packets, and sent to Charlie separately in quantum channels $X$ and $Y$,
respectively; thus single photon nonlocality is presented. This is done by
sending a single photon either from the source $S_{0}$ (sending $\left\vert
\Psi _{0}\right\rangle $) or from the source $S_{1}$ (sending $\left\vert
\Psi _{1}\right\rangle $), then splitting it with the beam splitter $BS_{1}$
made of a half-silvered mirror (note that polarizing beam splitters are not
recommended due to the security problem addressed at the end of section 6 of
\cite{HeJPA}).

To ensure the security of the transmission, the wave packet in channel $Y$
is delayed by the storage ring $SR_{1}$, which introduces a sufficiently
long delay time $\tau $ so that this wave packet will not leave Alice's site
until the other wave packet in channel $X$ already entered Charlie's site.
Thus the two wave packets of the same photon are never present together in
the transmission channels. This makes it impossible for Eve to prepare and
send Charlie a perfect clone \textit{on time} if she waits to intercept and
measure both wave packets, even though $\left\vert \Psi _{0}\right\rangle $\
and $\left\vert \Psi _{1}\right\rangle $ are orthogonal, because she has to
decide what to resend to channel $X$ before she can receive anything from
channel $Y$. On the other hand, when no eavesdropping occurs, Charlie can
distinguish $\left\vert \Psi _{0}\right\rangle $\ and $\left\vert \Psi
_{1}\right\rangle $ unambiguously by adding a storage ring $SR_{2}$ to
channel $X$ whose delay time is also $\tau $, while introducing a phase
shift $\pi $ to channel $Y$. The two wave packets of the same photon will
then recombine and interfere on the beam splitter $BS_{2}$, which is
identical to $BS_{1}$. Thus the complete apparatus of Alice and Charlie
forms a Mach-Zehnder interferometer, so that $\left\vert \Psi
_{0}\right\rangle $ ($\left\vert \Psi _{1}\right\rangle $) will always make
the detector $D_{0}$ ($D_{1}$) click with certainty, allowing Charlie to
decode the transmitted bit correctly. Any mismatched result between Alice's
transmitted state and Charlie's measurement will immediately reveals the
presence of Eve \cite{qi917}.

Comparing with the GV QKD scheme \cite{qi858}, the key difference is that $%
BS_{1}$\ and $BS_{2}$ in the KI scheme are asymmetric beam splitters, while
the GV scheme uses symmetric ones. The advantage of this modification is
that the sending time of each photon can be fixed and publicly known
beforehand, while in the GV scheme it has to be random and kept secret from
Eve until the security check.

An important fact that will be useful for our QBC protocol is that, since
the KI scheme is unconditionally secure, it is clear that Eve's
intercept-resend attack will always be detected with a nontrivial
probability. That is, if she intercepts a state sent from Alice and decodes
a nontrivial amount of information, then the state she resends to Charlie
cannot always make Charlie's correct detector click at the right time with
the probability $100\%$, no matter what is Eve's strategy on preparing the
resent state. There will always be a nontrivial probability that the resent
state will be detected by the wrong detector (the one that should not click
when no eavesdropping presented), or at the wrong time (which is earlier or
later than the correct arrival time when the state is not intercepted).
Please see \cite{qi917} for the rigorous proof, as well as some examples
showing why Eve's strategies do not work.

\section{Our QBC protocol}


\begin{figure*}[tbp]
\includegraphics{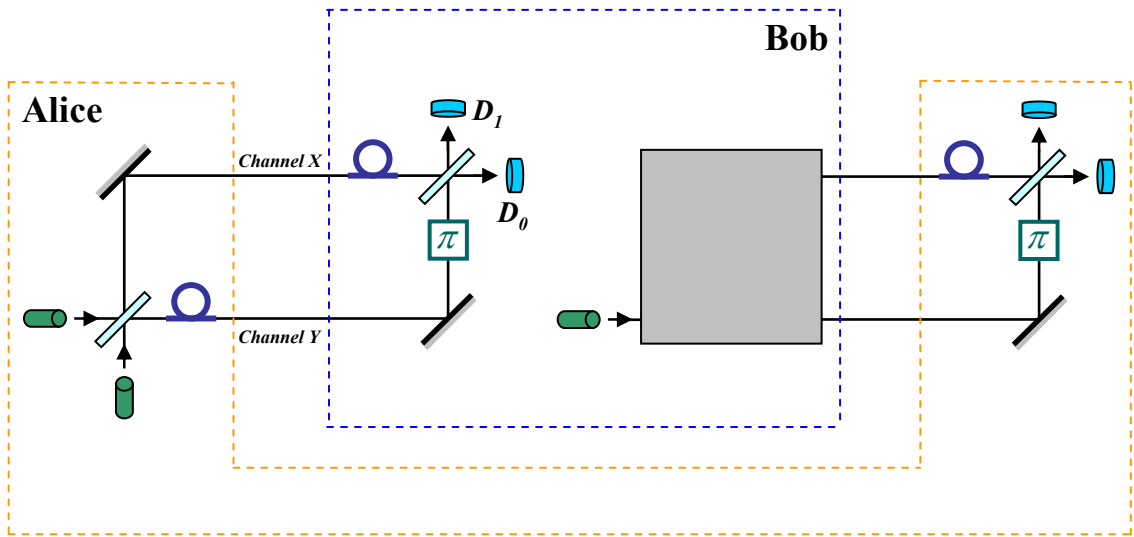}
\caption{Diagram for the apparatus of the QBC protocol when Bob chooses the
intercept mode. He measures Alice's photon using the same device as that of
Alice's, while sending another photon to Alice according to a certain
strategy (corresponding to the device illustrated as the black box in the
diagram) to reduce Alice's probability of finding his interception.}
\label{fig:epsart}
\end{figure*}


As illustrated in FIG. 2, to build a QBC protocol upon the above KI QKD
scheme, we treat Charlie's site as a part of Alice's, so that the two
parties merge into one. Alice sends out a bit-string $c$ encoded with the
above orthogonal states, whose value is related to the bit $b$ she wants to
commit. Then she receives the states herself. Meanwhile, let Bob take the
role of Eve. His action shifts between two modes. In the \textit{bypass}
mode, he simply does nothing so that the corresponding parts of the states
return to Alice intact. In the \textit{intercept} mode, he applies the
intercept-resend attack. That is, he intercepts the state and decodes the
corresponding bit (which can be done using the same device as that of
Charlie's), and prepares a fake state to resend to Alice on time. While
there could be many strategies for Bob to prepare the resent state, we use $%
\varepsilon $ to denote the lower bound of the average probability for his
resent state to be caught in Alice's check. As we mentioned above, the
unconditional security of the KI QKD scheme guarantees that $\varepsilon $
cannot always equal exactly to zero for both $\left\vert \Psi
_{0}\right\rangle $\ and $\left\vert \Psi _{1}\right\rangle $ even when Bob
uses the optimal strategy. Therefore, Alice is able to estimate the upper
bound of the frequency of the presence of the intercept mode, to limit Bob
from intercepting the whole string $c$, so that the value of the committed
bit $b$ can be made concealing. Meanwhile, since $\varepsilon <1$, at the
end of the commit phase there will be some bits of the string become known
to Bob, while Alice does not know the exact position of all these bits. Thus
she cannot alter string $c$ freely at a later time, making the protocol
binding. The complete QBC protocol is described below.

\bigskip

The \textit{commit} protocol:

(1) Bob chooses a binary linear $(n,k,d)$-code $C$ \cite{qi43} and announces
it to Alice, where $n$, $k$, $d$ are agreed on by both Alice and Bob.

(2) Alice chooses a nonzero random $n$-bit string $r=(r_{1}r_{2}...r_{n})\in
\{0,1\}^{n}$ and announces it to Bob. This makes any $n$-bit codeword $%
c=(c_{1}c_{2}...c_{n})$ in $C$ sorted into either of the two subsets $%
C_{(0)}\equiv \{c\in C|c\odot r=0\}$ and $C_{(1)}\equiv \{c\in C|c\odot
r=1\} $. Here $c\odot r\equiv \bigoplus\limits_{i=1}^{n}c_{i}\wedge r_{i}$ .

(3) Now Alice decides the value of the bit $b$ that she wants to commit.
Then she chooses a codeword $c$ from $C_{(b)}$ randomly.

(4) Alice encodes each bit of this specific $c$ as $c_{i}\rightarrow
\left\vert \Psi _{c_{i}}\right\rangle $ where the state $\left\vert \Psi
_{c_{i}}\right\rangle $ is defined by equation (\ref{eqpsi}). She sends Bob
the two wave packets of the same state separately in channels $X$ and $Y$,
with the storage ring $SR_{1}$ on channel $Y$ introducing a delay time $\tau
$ known to Bob.

(5) For each of Alice' states $\left\vert \Psi _{c_{i}}\right\rangle $ ($%
i=1,...,n$), Bob chooses the intercept mode with probability $f$ ($f<1-d/n$)
and the bypass mode with probability $1-f$.

If Bob chooses to apply the bypass mode, he simply keeps channels $X$ and $Y$
intact so that the state sent from Alice will be returned to her detectors
as is.

Else if Bob chooses to apply the intercept mode, he uses the same
measurement device as that of Alice's, to measure $\left\vert \Psi
_{c_{i}}\right\rangle $ so that he can decode $c_{i}$\ with certainty.
Meanwhile, he prepares another state and sends it back to channels $X$ and $%
Y $ at the right time, i.e., the time which can ensure that it reaches
Alice's detectors at a time that looks as if Bob were applying the bypass
mode.

There are many different strategies how Bob sends this state (thus we left
this part of Bob's device as a black box in FIG. 2). One of the simplest
ways is to use the same device that Alice uses for sending her state. More
strategies will be discussed in section V.A. It will be shown there that all
these strategies cannot hurt the security of the protocol, so that Bob are
allowed to use any of them here. As stated above, in any strategy there is a
nonvanishing probability that Bob's resent state does not equal to Alice's
original $\left\vert \Psi _{c_{i}}\right\rangle $ since he has to send it
before completing the measurement on $\left\vert \Psi _{c_{i}}\right\rangle $%
\ (to be further explained in section V.A too). Let $\varepsilon $ be the
lower bound of this probability for all these strategies.

(6) Alice uses the same device that Charlie used in the KI QKD scheme, to
measure the output of the quantum channels from Bob. She counts how many
times her measurement results do not match the states she sent, and denotes
it as $n^{\prime }$. From step (5) it can be shown that $n^{\prime }\sim
\varepsilon fn$. Thus Alice can estimate the upper bound of the probability
of Bob choosing the intercept mode as $f\sim n^{\prime }/(\varepsilon n)$.
Alice agrees to continue with the protocol if she finds $f<1-d/n$, which
means that the number of $c_{i}$'s known to Bob is $fn<n-d$. Otherwise she
concludes that Bob is cheating.

\bigskip

The \textit{unveil} protocol:

(7) Alice announces the values of $b$ and the specific $%
c=(c_{1}c_{2}...c_{n})$ she chose.

(8) Bob accepts the commitment if $c\odot r=b$ and $c$ is indeed a codeword
from $C$, and every $c_{i}$\ agrees with the state $\left\vert \Psi
_{c_{i}}\right\rangle $ he detected in the intercept mode.

\section{Security}

\subsection{Basic ideas}

For easy understanding, here we will first give a heuristic explanation of
the security of the protocol. A more general theoretical proof will be
provided in section V.B.

The binary linear $(n,k,d)$-code $C$ used in the protocol can simply be
viewed as a set of classical $n$-bit strings. Each string is called a
codeword. This set of strings has two features.

(A) Among all the $2^{n}$ possible choices of $n$-bit strings, only a
particular set of the size $\sim 2^{k}$\ ($k<n$) is selected to form this
set.

(B) The distance (i.e., the number of different bits) between any two
codewords in this set is not less than $d$\ ($d<n$).

Feature (A) puts a limit on Alice's freedom on choosing the initial state $%
\left\vert \psi _{c}\right\rangle \equiv \left\vert \Psi
_{c_{1}}\right\rangle \otimes \left\vert \Psi _{c_{2}}\right\rangle \otimes
...\otimes \left\vert \Psi _{c_{i}}\right\rangle \otimes ...\otimes
\left\vert \Psi _{c_{n}}\right\rangle $, since each $\left\vert \Psi
_{c_{i}}\right\rangle $ ($i=1,...,n$) cannot be chosen independently.
Instead, the string $c=(c_{1}c_{2}...c_{n})$ formed by the indices $c_{i}$'s
can only be chosen among the codewords. Meanwhile, feature (B) guarantees
that if Alice wants to change the string $c$ from one codeword into another
so that the value of her committed $b$ can be altered, she needs to change
at least $d$ bits of the codeword $c$. But among the $n$ states she sent to
Bob, there are only $n^{\prime }\sim \varepsilon fn$ states which she knows
for sure that Bob has applied the intercept mode. For each of the rest $%
n-n^{\prime }$ states, her measurement result always matches the state she
sent, so that she cannot distinguish unambiguously whether Bob has applied
the bypass mode or the intercept mode. If it was the intercept mode, Bob
already knew the corresponding bit $c_{i}$ of the codeword $c$, so that
Alice's altering it will be caught inevitably. Rigorously speaking, among
these $n-n^{\prime }$ states, the probability that Bob has applied the
intercept mode is%
\begin{equation}
p=\frac{fn-n^{\prime }}{n-n^{\prime }}=\frac{f-\varepsilon f}{1-\varepsilon f%
}.
\end{equation}%
Therefore Alice's altering one bit of the codeword will stand the
probability $p$ to be detected, and her probability of altering $\geq d$
bits without being detected will be $(1-p)^{d}$. Alternatively, Alice may
prepare $\left\vert \psi _{c}\right\rangle $ initially in a state where $c$
is not a codeword. Instead, it is half-way between two codewords, so that
changing $d/2$ will be sufficient to turn it into one of the codewords. In
this case, her probability of escaping the detection will be increased to $%
(1-p)^{d/2}$. Nevertheless, in either way the probability drops
exponentially as $d$ increases. Thus it can be made arbitrarily close to
zero.

On the other hand, feature (A) also guarantees that the number of different
codewords having less than $n-d$ bits in common increases exponentially with
$k$. This is the key that makes our protocol secure against Bob, as Alice
can bound the frequency $f$ that Bob applies the intercept mode, which in
turns bounds the number of the bits known to Bob so that he cannot know
Alice's actual choice of the codeword. The reason is that in step (5) of the
protocol, no matter what is the strategy that Bob used in his intercept mode
to prepare the resent state, it can be shown that his resent state cannot
\textit{always} equal to the $\left\vert \Psi _{c_{i}}\right\rangle $ he
received from Alice, as long as he wants to ensure that the time it reaches
Alice's detectors will show no difference than the case where he were
applying the bypass mode. That is, either Bob's resent state in the
intercept mode will inevitably make Alice's detectors click earlier or later
than the time it does when the bypass mode were used instead, or the resent
state will be different from $\left\vert \Psi _{c_{i}}\right\rangle $ with a
nonvanishing probability $\varepsilon $.

To see why this is the case, let us first consider the strategy where Bob
prepares the resent state using the same device that Alice uses for sending
her $\left\vert \Psi _{c_{i}}\right\rangle $. Suppose that the first wave
packet of Alice's $\left\vert \Psi _{c_{i}}\right\rangle $ enters Bob's site
via channel $X$ (before entering his storage ring) at time $t_{i}$ (which
should be agreed on by both Alice and Bob beforehand). Then the second wave
packet of $\left\vert \Psi _{c_{i}}\right\rangle $ will enter Bob's site via
channel $Y$ at time $t_{i}+\tau $ due to the existence of the storage ring
in Alice's sending device (see FIG. 2). As there is also another storage
ring in Alice's measurment device, Bob must send the first wave packet of
his resent state into channel $X$ at time $t_{i}$, and send the second wave
packet into channel $Y$ at time $t_{i}+\tau $, so that they can reach
Alice's measurement device at a time that looks like Bob is running the
bypass mode. Here for simplicity, we assume that the time for the wave
packets to travel the rest part of the channels (other than the storage
rings) is negligible. That is, the first wave packet of Bob's state needs to
be sent before the second wave packet of Alice's $\left\vert \Psi
_{c_{i}}\right\rangle $ reaches him. Otherwise his resent state will reach
Alice's detectors later than the expected arrival time when no interception
occurs. Therefore, by the time Bob prepares the resent state, he has not
received Alice' state completely yet so he does not know the form of $%
\left\vert \Psi _{c_{i}}\right\rangle $. Thus his resent state cannot match
Alice's state with probability $100\%$, and he cannot make them match by his
local operations acting solely on the second wave packet of his resent state
(the one to be sent via channel $Y$) after the first wave packet in channel $%
X$ was already sent.

On the other hand, suppose that Bob waits until $t_{i}+\tau $ so that
Alice's $\left\vert \Psi _{c_{i}}\right\rangle $ enters his measurement
device completely, and prepares the resent state following the measurement
result. Then even though the form of the states will match perfectly, the
resent state will reach Alice's detectors much later than it should when Bob
uses the bypass mode, because the storage ring in channel $X$ of Alice's
measurement device will delay the corresponding wave packet. This will make
Alice aware that Bob is using the intercept mode too.

Alternatively, there can be another strategy where Bob simply sends all wave
packets of his state simultaneously via one of the channels alone, e.g.,
through channel $X$ at time $t_{i}$ or through channel $Y$ at time $%
t_{i}+\tau $. 
But he will not be able to guarantee which of Alice's detectors will click
with certainty.

Though here we have only analyzed the above few strategies as examples, the
result is common. That is, in any strategy potentially existed, there will
be a nonvanishing probability (let $\varepsilon $ denote the lower bound of
the probability for all strategies) that Alice will find a mismatched result
between the $\left\vert \Psi _{c_{i}}\right\rangle $ she sent and her
measurement on the state she received from Bob, or the arrival time of Bob's
state is different from what it should be in the bypass mode. This is
because, as mentioned in section IV, the role of Bob in our protocol is
actually the same as that of Eve in the KI QKD scheme \cite{qi917}. If there
is a strategy which is not bounded by the above result, then it can be
utilized to fulfill a successful eavesdropping to the KI QKD scheme, which
will conflict with the existing proof of the unconditional security of the
scheme \cite{qi917}.

As a consequence, whenever Bob applies the intercept mode, Alice can
distinguish it from the bypass mode with the probability $\varepsilon $.
Then as described in step (6), by counting the number $n^{\prime }$ of the
mismatched results, Alice can estimate the upper bound of the probability of
Bob choosing the intercept mode as $f\sim n^{\prime }/(\varepsilon n)$. As
she agrees to continue only when $f<1-d/n$, it is guaranteed that during the
commit phase, Bob knows only $fn<n-d$ bits of $c$. Since feature (A) of the
binary linear $(n,k,d)$-code $C$ ensures that the number of different
codewords having $fn$ bits in common increases exponentially with $k$, the
potential choices for $c$ are too much for Bob to determine whether $c$
belongs to the subset $C_{(0)}$ or $C_{(1)}$. Thus the amount of information
Bob gained on the committed bit $b$ before the unveil phase can be made
arbitrarily close to zero by increasing $k$.

Taking both Alice's and Bob's cases studied above into consideration, we can
see that fixing $k/n$ and $d/n$ while increasing $n$ will then result in a
protocol secure against both sides.

\subsection{More general proof: evading the MLC no-go theorem}

While it is hard to find a completely general proof like those for QKD \cite%
{qi70} showing that the protocol is secure against all cheating strategies
that may potentially exist, here it can be shown that our protocol is at
least secure against the specific cheating strategy proposed in the MLC
theorem.

\subsubsection{Essential of the no-go proof}

As pinpointed out in section 4 of \cite{HeJPA}, while there are many
variations and extensions of the MLC no-go theorem \cite{qi74}-\cite{qbc32},
their proofs all have the following common features.

(I) The reduced model. According to the no-go proofs, any QBC protocol can
be reduced to the following model. Alice and Bob share a quantum state in a
given Hilbert space. Each of them performs unitary transformations on the
state in turns. All measurements are performed at the very end.

(II)\ The coding method. The quantum state corresponding to the committed
bit $b$ has the form%
\begin{equation}
\left\vert \psi _{b}\right\rangle =\sum\limits_{j}\lambda
_{j}^{(b)}\left\vert e_{j}^{(b)}\right\rangle _{A}\otimes \left\vert
f_{j}^{(b)}\right\rangle _{B}.  \label{coding}
\end{equation}%
Here the systems $A$ and $B$ are owned by Alice and Bob respectively.

(III) The concealing condition. To ensure that Bob's information on the
committed bit is trivial before the unveil phase, any QBC protocol secure
against Bob should satisfy%
\begin{equation}
\rho _{0}^{B}\simeq \rho _{1}^{B},  \label{eqconcealing}
\end{equation}%
where $\rho _{b}^{B}\equiv Tr_{A}\left\vert \psi _{b}\right\rangle
\left\langle \psi _{b}\right\vert $ is the reduced density matrix
corresponding to Alice's committed bit $b$, obtained by tracing over system $%
A$ in equation (\ref{coding}). Note that in some presentation of the no-go
proofs (e.g., \cite{qi147,qbc31,qi323,qi714}), this feature was expressed
using the trace distance or the fidelity instead of the reduced density
matrices, while the meaning remains the same.

(IV) The cheating strategy. As long as equation (\ref{eqconcealing}) is
satisfied, according to the Hughston-Jozsa-Wootters (HJW) theorem (a.k.a.
the Uhlmann theorem, etc.) \cite{qi73}, there exists a local unitary
transformation for Alice to map $\{\left\vert e_{j}^{(0)}\right\rangle
_{A}\} $ into $\{\left\vert e_{j}^{(1)}\right\rangle _{A}\}$ successfully
with a high probability \cite{qi73}. Thus a dishonest Alice can unveil the
state as either $\left\vert \psi _{0}\right\rangle $ or $\left\vert \psi
_{1}\right\rangle $\ at her will with a high probability to escape Bob's
detection. Consequently, a concealing QBC protocol cannot be binding.

\subsubsection{Equivalent model of our protocol}

Following the method of the MLC theorem to reduce our protocol into an
equivalent form where Alice and Bob perform unitary transformations in turns
on the quantum state they share, we can see that our commit protocol is
essentially the following 3-stage process.

(i) Alice prepares and sends Bob a system $\alpha $ containing $n$ qubits $%
\left\vert \Psi _{c_{i}}\right\rangle $ ($i=1,...,n$).

(ii) Bob prepares an $n$-qubit system $\beta $ and an $n$-qutrit system $%
\gamma $. Then he performs an operation $U_{B}$ on the combined system $%
\beta \otimes \alpha \otimes \gamma $.

(iii) Bob sends system $\beta $ to Alice.

In this process, the initial state of each qubit in $\beta $ is chosen at
Bob's choice, and kept secret from Alice. The system $\gamma $ is for
storing the result of Bob's measurement on $\alpha $. The three orthogonal
states of each qutrit in $\gamma $ are denoted as $\left\vert 0\right\rangle
$, $\left\vert 1\right\rangle $\ and $\left\vert 2\right\rangle $, where $%
\left\vert 0\right\rangle $ ($\left\vert 1\right\rangle $) means that the
measurement result of $\left\vert \Psi _{c_{i}}\right\rangle $ is $c_{i}=0$ (%
$c_{i}=1$), and $\left\vert 2\right\rangle $\ means that $\left\vert \Psi
_{c_{i}}\right\rangle $ is not measured so that the result remains unknown.
All the $n$ qutrits in system $\gamma $ are initialized in $\left\vert
2\right\rangle $. The operation $U_{B}=u_{1}\otimes u_{2}\otimes ...\otimes
u_{n}$ is defined by the mode that Bob chooses for each qubit in $\alpha $,
where each $u_{i}$ ($i=1,...,n$) is a $3$-particle operator acting on the $i$%
-th qubits/qutrit of $\beta $, $\alpha $ and $\gamma $. If Bob chooses the
bypass mode for the $i$-th qubit of $\alpha $, then%
\begin{equation}
u_{i}=u_{byp}\equiv P_{\beta \alpha }\otimes I_{\gamma }.  \label{u bypass}
\end{equation}%
Here $P_{\beta \alpha }$ is a $2$-qubit permutation operator which
interchanges the states of the $i$-th qubits of $\alpha $ and $\beta $, and $%
I_{\gamma }$ is the identity operator that keeps the $i$-th qutrit of $%
\gamma $\ unchanged. Or if Bob chooses the intercept mode for the $i$-th
qubit of $\alpha $, then%
\begin{equation}
u_{i}=u_{int}\equiv I_{\beta }\otimes (\left\vert \Psi _{0}\right\rangle
_{\alpha }\left\langle \Psi _{0}\right\vert \otimes \mu _{\gamma
}^{(0)}+\left\vert \Psi _{1}\right\rangle _{\alpha }\left\langle \Psi
_{1}\right\vert \otimes \mu _{\gamma }^{(1)}).  \label{u intercept}
\end{equation}%
Here $I_{\beta }$ is the identity operator on the $i$-th qubit of $\beta $, $%
\mu _{\gamma }^{(0)}$ ($\mu _{\gamma }^{(1)}$) is an unitary transformation
on the $i$-th qutrit of $\gamma $ which can map the state $\left\vert
2\right\rangle $\ into $\left\vert 0\right\rangle $\ ($\left\vert
1\right\rangle $). That is, applying $u_{int}$\ is equivalent to measuring
the $i$-th qubit of $\alpha $ and storing the result in the $i$-th qutrit of
$\gamma $, while keeping the $i$-th qubit of $\beta $ unchanged. Among all
the possible forms of the operation $U_{B}$, Bob chooses one of those that
can pass the security check in step (6) of our original protocol in section
IV (i.e., the number of $u_{int}$ in $U_{B}$\ should be $fn<n-d$), and keeps
his choice secret from Alice.

\subsubsection{Security against Alice's cheating}

Now we will show why the specific cheating strategy in the MLC theorem does
not apply to our protocol. With the above equivalent description, we can see
that after the commit phase, the density matrix of the systems shared
between Alice and Bob is in the form $U_{B}(\rho ^{\beta }\otimes \rho
^{\alpha }\otimes \rho ^{\gamma })U_{B}^{\dagger }$, where $\rho ^{\beta }$,
$\rho ^{\alpha }$ and $\rho ^{\gamma }$ are the density matrices for the
systems $\beta $, $\alpha $ and $\gamma $, respectively. Note that here the
first $n$ qubits are now owned by Alice, i.e., they serve like the system $A$
in equation (\ref{coding}) even though system $\beta $\ was prepared by Bob.
Meanwhile, the rest qubits and qutrits now belong to Bob and serve like the
system $B$ in equation (\ref{coding}), even though system $\alpha $\ was
prepared by Alice.

Let $\rho _{0}^{\alpha }$ ($\rho _{1}^{\alpha }$) denote the initial density
matrix of system $\alpha $\ prepared by Alice, that can unveil the committed
bit as $b=0$ ($b=1$). According to the MLC theorem, now the goal for an
dishonest Alice is to find a local unitary transformation $U_{A}$ acting on
the $n$ qubits at her side alone, that can map $\rho _{0}^{B}$ into $\rho
_{1}^{B}$. That is, $U_{A}$ should satisfy%
\begin{equation}
U_{A}[U_{B}(\rho ^{\beta }\otimes \rho _{0}^{\alpha }\otimes \rho ^{\gamma
})U_{B}^{\dagger }]U_{A}^{\dagger }=U_{B}(\rho ^{\beta }\otimes \rho
_{1}^{\alpha }\otimes \rho ^{\gamma })U_{B}^{\dagger }.
\end{equation}%
Applying $U_{B}^{\dagger }(...)U_{B}$\ on both sides of this equation, we
yield%
\begin{equation}
(U_{B}^{\dagger }U_{A}U_{B})(\rho ^{\beta }\otimes \rho _{0}^{\alpha
}\otimes \rho ^{\gamma })(U_{B}^{\dagger }U_{A}U_{B})^{\dagger }=\rho
^{\beta }\otimes \rho _{1}^{\alpha }\otimes \rho ^{\gamma }.  \label{UA}
\end{equation}%
Recall that $U_{B}$ was kept secret from Alice, thus she has to choose an $%
U_{A}$ which is independent of $U_{B}$. Meanwhile, the right side of
equation (\ref{UA}) is independent of $U_{B}$ too. Therefore, equation (\ref%
{UA}) cannot be satisfied in general, unless $U_{A}$ always commutes with
the specific $U_{B}$'s that Bob may choose in the protocol. In this case $%
U_{B}^{\dagger }U_{A}U_{B}=U_{A}U_{B}^{\dagger }U_{B}=U_{A}$, and equation (%
\ref{UA}) becomes%
\begin{equation}
U_{A}(\rho ^{\beta }\otimes \rho _{0}^{\alpha }\otimes \rho ^{\gamma
})U_{A}^{\dagger }=\rho ^{\beta }\otimes \rho _{1}^{\alpha }\otimes \rho
^{\gamma }.  \label{UA2}
\end{equation}%
In many previous QBC protocols (e.g. \cite{qi365,qi43}) there is $\rho
_{0}^{\alpha }=\rho _{1}^{\alpha }$. Then the HJW theorem guarantees that
such a local $U_{A}$ exists, so that Alice can alter the value of her
committed bit. This is why the cheating strategy in the MLC theorem is
always successful in such protocols. However, as shown from equation (\ref%
{eqpsi}), in our protocol $\left\vert \Psi _{0}\right\rangle $ and $%
\left\vert \Psi _{1}\right\rangle $ are orthogonal. Thus the state $%
\left\vert \psi _{c}\right\rangle \equiv \left\vert \Psi
_{c_{1}}\right\rangle \otimes \left\vert \Psi _{c_{2}}\right\rangle \otimes
...\otimes \left\vert \Psi _{c_{i}}\right\rangle \otimes ...\otimes
\left\vert \Psi _{c_{n}}\right\rangle $ corresponding to a specific codeword
$c$ is orthogonal to the state corresponding to any other codeword.
Consequently, our QBC protocol satisfies $\rho _{0}^{\alpha }\perp \rho
_{1}^{\alpha }$, which is a crucial difference from previous insecure
protocols. In this case the HJW theorem does not apply, making Alice unable
to find an local transformation $U_{A}$ acting on her system (i.e., system $%
\beta $) alone while satisfying equation (\ref{UA2}). For this reason, our
protocol is immune to the specific cheating strategy proposed in the MLC
no-go theorem.

\subsubsection{Security against Bob's cheating}

As the system $\alpha $\ Alice sent to Bob in our QBC protocol satisfies $%
\rho _{0}^{\alpha }\perp \rho _{1}^{\alpha }$, at the first glance it seems
that Bob can simply perform a measurement $M$\ to project the state of $%
\alpha $ into the Hilbert spaces supported by $\rho _{b}^{\alpha }$ ($b=0,1$%
)\ and thus learn the value of Alice's committed $b$. But this is not true,
because as shown above, the density matrix of the systems shared between
Alice and Bob after the commit phase is $U_{B}(\rho ^{\beta }\otimes \rho
^{\alpha }\otimes \rho ^{\gamma })U_{B}^{\dagger }$. That is, Bob is
required to perform the operation $U_{B}$ on system $\alpha $. This $U_{B}$
is incommutable with the measurement $M$\ on $\alpha $\ for distinguishing $%
\rho _{0}^{\alpha }$ and $\rho _{1}^{\alpha }$, since equation (\ref{u
bypass}) indicates that $U_{B}$ contains permutation operators which act on
not only system $\alpha $, but also other systems. Therefore, $U_{B}$ and $M$%
\ cannot be both applied on $\alpha $. Bob can only choose one of them. The
timing of the sending of the states in the protocol also prevents Bob from
keeping system $\beta $ unsent until he receives the entire system $\alpha $%
. Thus he needs to decide on the fly which operation to apply. Once he
applied $U_{B}$, then the state of $\alpha $ will be disturbed so that Bob
will lose the chance to apply $M$ on it for decoding $b$.

Moreover, the difference between $U_{B}$ and $M$\ is detectable to Alice.
According to step (5) of our protocol, an $U_{B}$ is considered legitimate
if it includes the operator $u_{byp}$ (i.e., Bob is using the bypass mode)
for $(1-f)n>d$\ times. On the contrary, because the minimal distance between
the codewords is $d$, the number of $u_{byp}$ in $M$ must be less than $d$.
Otherwise as a basic property of the binary linear $(n,k,d)$-code, the
number of the possible codewords having less than $n-d$ bits in common will
be at the order of magnitude of $2^{k}$ so that such an $M$ can provide only
trivial information on the value of Alice's committed $b$. But as elaborated
in section V.A, the unconditional security of the KI QKD scheme \cite{qi917}
ensures that, for each of Alice's state $\left\vert \Psi
_{c_{i}}\right\rangle $ which Bob applies the intercept mode, his resent
state can equal to $\left\vert \Psi _{c_{i}}\right\rangle $ with the
probability $1-\varepsilon $ only, where there is always $\varepsilon >0$ no
matter which resend strategy Bob uses. When the number of the applied bypass
mode is less than $d$, Alice will find that the number of mismatched result
between her received state and the original $\left\vert \Psi
_{c_{i}}\right\rangle $ she sent will be $n^{\prime }>\varepsilon (n-d)$. If
she takes $f\equiv n^{\prime }/(\varepsilon n)$ as stated in step (6) of the
protocol, then she will find that $f>1-d/n$. Thus she knows that Bob was
attempting to apply the operation $M$ instead of a legitimate $U_{B}$.

Namely, while Alice's committed state satisfies $\rho _{0}^{\alpha }\perp
\rho _{1}^{\alpha }$ so that it is distinguishable to Bob, Alice's security
check in the protocol requires Bob to perform an operation $U_{B}$, which is
incommutable with the operation $M$ for distinguishing $\rho _{0}^{\alpha }$
and $\rho _{1}^{\alpha }$. The protocol thus becomes concealing against Bob.

\section{Feasibility}

In the above we focused only on the theoretical possibility of evading the
MLC no-go theorem. But we can see from FIG. 2 that our protocol is also very
feasible, as the required experimental technology is already available today
\cite{qi889}.

Also, as the secret random sending time is no longer required, the commit
phase will take less time than that of our previous protocol \cite{HeJPA},
and the total number of photons that Bob needs to send in the intercept mode
will be significantly reduced. More rigorously, suppose that the minimal
time for Bob to shift between the intercept and bypass modes is $\Delta $.
When both protocols choose the same $n$ as the length of the codewords, our
previous protocol \cite{HeJPA} requires Bob to send $sf$\ ($s>>n$. Note that
$f$ was denoted as $\alpha $ in \cite{HeJPA}) photons in total, and the
duration time of the commit phase is $s\Delta $. But in the current
protocol, the photon number is reduced to $nf$, and the duration time is $%
n\Delta $. As it was suggested in section 3 of \cite{HeJPA} that a typical
choice of the parameters is $s/n=10$, we can see that the current protocol
can be $10$ times more efficient than our previous one \cite{HeJPA}.

There is another advantage of removing the need to keep the sending time
secret at first. That is, now Alice and Bob can determine the binary linear $%
(n,k,d)$-code $C$ and decide on the sending time of the states $\left\vert
\Psi _{c_{i}}\right\rangle $ ($i=1,...,n$) beforehand, so that no more
classical communication is needed at all during the commit phase, unless one
of them finds the other participant cheating and announces to abort.
Therefore, it not only reduces the communication cost significantly, but
also makes it easier for security analysis and comparison with the MLC
theorem, which was generally deduced in a theoretical model where classical
communication is not presented directly.

Nevertheless, under practical settings, some more security checks should be
added against technical attacks. Especially, the physical systems
implementing the states may actually have other degrees of freedom, which
leave rooms for Alice's cheating. For example, she may send photons with
certain polarization or frequency, so that she can distinguish them from the
photons Bob sends in the intercept mode. In this case, Bob and Alice should
discuss at the beginning of the protocol, to limit these degrees of freedom
to a single mode. In step (5) when Bob chooses the intercept mode, he should
also measure occasionally these degrees of freedom of some of Alice's
photons, instead of performing the measurement in the original step (5).
Then if Alice wants to send distinguishable photons with a high probability
so that they are sufficient for her cheating, she will inevitably be
detected.

Moreover, when Bob applies the bypass mode, he should add phase shifters to
both channels $X$ and $Y$ to introduce the same phase shift in both channels
so that an honest Alice will not be affected (to be explained in the
appendix), while the amount of this phase shift is randomly chosen and kept
secret from Alice, so that the counterfactual attack described in the
appendix can be defeated.

Note that all these technical attacks and the corresponding modifications
are due to the imperfection of the physical systems implementing the
protocol. They should not be messed with the theoretical possibility of
evading the MLC no-go theorem.

\section{Summary and discussions}

In conclusion, inspired by the KI simplified version \cite{qi917} of the GV
QKD scheme \cite{qi858}, we improved our previously proposed QBC protocol
\cite{HeJPA}, and proved that it is immune to the specific cheating strategy
used in the MLC no-go theorem of unconditionally secure QBC. The key reason
is that the density matrices of Alice's states corresponding to the
committed values $0$ and $1$ are orthogonal to each other, making her unable
to find the local unitary transformation for the cheating by using the HJW
theorem. Meanwhile, the protocol remains concealing against Bob because he
is required to perform another operation, which is incommutable with the
measurement for distinguishing the density matrices.

However, there may potentially exist innumerous cheating strategies other
than the one in the MLC theorem. It is natural to ask whether our protocol
can be unconditionally secure against all these strategies. A rigorous
evaluation of the upper bound of the cheating probability is also needed.
For example, our protocol involves a probability $\varepsilon $ which
denotes the lower bound of the average probability for Bob's resent state in
step (5) to be caught in Alice's check. The exact value of $\varepsilon $
should be determined by the rigorous quantitatively security analysis of the
KI QKD scheme. Unfortunately, such a value was not yet provided in the
literature \cite{qi917}. In turns, the parameters $k$ and $d$\ in the binary
linear $(n,k,d)$-code $C$ used in our protocol need to be chosen according
to $\varepsilon $. Therefore it remains unknown whether the cheater can have
a nontrivial probability of success if these parameters are improperly
chosen. We would like to leave these questions open for future researches.

Though the current QBC protocol and the one in \cite{HeJPA} have
similarities in many ways, the underlying origins of their security against
Bob are somewhat different. While both protocols are immune to Bob's
cheating because they are based on unconditionally secure QKD schemes, as
pointed out in \cite{qi917}, the GV QKD scheme can actually be viewed as
utilizing three orthogonal states -- two photon states and one vacuum state.
Its security is provided by the random sending time of the photons. On the
contrary, the KI QKD scheme does not require the vacuum state and the secret
sending time. The security is guaranteed by the fact that the eavesdropper
cannot fake the states with certainty owe to the use of the asymmetric beam
splitters. Similarly, the security of the QBC protocol in \cite{HeJPA}
against Bob is based on Alice's random sending time that remains secret
before the last step of the commit phase, while in our current QBC proposal,
it is because Bob cannot fake the states with certainty when he runs the
intercept mode. Therefore our current protocol is more than merely a
simplification on the presentation.

The work was supported in part by the NSF of China,
the NSF of Guangdong province, and the Foundation of Zhongshan University
Advanced Research Center.

\appendix

\section{Defeating the counterfactual attack}

As we mentioned in Sec. 6, under practical settings our QBC protocol may
need some modifications against technical attacks. Here we give such an
example.

Recently a cheating strategy against counterfactual QKD protocols \cite%
{qi801,qi1026} was proposed \cite{qi1025}. Unlike general intercept-resend
attacks in which measurements are performed on the quantum states carrying
the secret information, in this strategy the cheater makes use of quantum
counterfactual effect to detect the working modes of the devices of other
participants. Thus it was named \textquotedblleft the counterfactual
attack\textquotedblright\ \cite{qi1025}. Here we will skip how it applies to
QKD protocols, while focus only on its impact on our QBC protocol.


\begin{figure*}[tbp]
\includegraphics{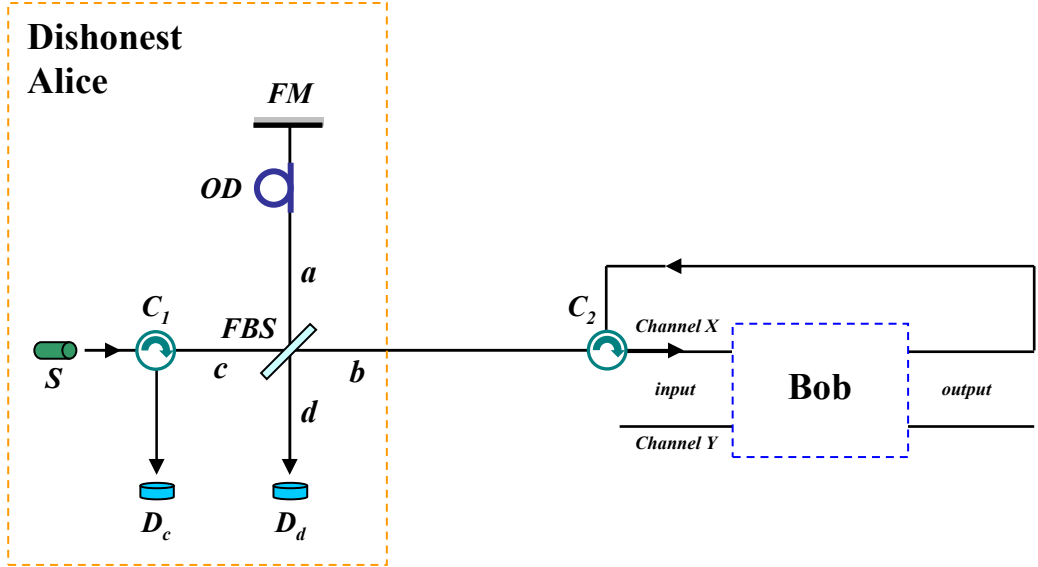}
\caption{Diagram of the apparatus for Alice's counterfactual attack. A
single-photon pulse produced by the source $S$ passes through the optical
circulator $C_{1}$ and hits the \textquotedblleft
fictitious\textquotedblright\ beam splitter (FBS) along path $c$. Path $a$
is adjusted by the optical delay $OD$, followed by a Faraday mirror $FM$.
Any photon coming from path $c$ from the right to the left will be detected
by the detector $D_{c}$, while the detector $D_{d}$ detects any photon
coming from path $d$. Path $b$ is connected to both the input and output of
Bob's channel $X$ (or both the input and output of Bob's channel $Y$) via
the optical circulator $C_{2}$.}
\label{fig:epsart}
\end{figure*}


FIG. 3 illustrates the apparatus for the attack \cite{qi1025}. The core is a
\textquotedblleft fictitious\textquotedblright\ beam splitter (FBS) which
has the following functions.

(f1) Any photon hitting the FBS from path $c$\ will be reflected with
certainty.

(f2) When the paths $a$ and $b$ are adjusted correctly, two wave packets
coming from paths $a$ and $b$, respectively, will interfere and combine
together, and enter path $c$ with certainty.

(f3) Any photon hitting the FBS from path $a$ alone will pass through the
FBS and enter path $d$ with certainty.

An ideal FBS that can realize these functions faithfully does not exist in
principle. Thus it is called \textquotedblleft fictitious\textquotedblright
. For example, devices with the functions (f2) and (f3) may not accomplish
the function (f1) perfectly, i.e., a photon coming from path $c$ could pass
the devices with a nontrivial probability, making the attack detectable.
However, FBS can be implemented approximately by using an infinite number of
ordinary BS \cite{qi1026,qi1025}. In practice, the number of BS involved in
the implementation has to be finite. But if the deviation from an ideal FBS
is too small to be detected within the capability of available technology,
then the attack could become a real threat.

Suppose that an ideal FBS is available to a dishonest Alice in our QBC
protocol. When she is supposed to send Bob a state\ in step (4), she runs
both the FBS system in FIG. 3 and the apparatus in the honest protocol
(i.e., the one shown in FIG. 2) simultaneously in parallel, with path $b$ of
the FBS system connecting to both the input and output of Bob's channel $X$
(or both the input and output of Bob's channel $Y$). The apparatus in FIG. 2
works as usual so that the protocol can be executed as if she is honest,
while the FBS system serves as a probe to detect Bob's mode. According to
the function (f2) of the FBS, whenever Bob applies the bypass mode in step
(5), the wave packets of a photon Alice sent to the FBS will be returned
from both paths $a$ and $b$ so that the detector $D_{c}$ will click with
certainty. On the other hand, whenever Bob applies the intercept mode, an
ideal FBS can guarantee that $D_{c}$ will never click as path $b$ is
actually blocked. Therefore Alice can learn Bob's mode unambiguously. Since
Bob does not know the state $\left\vert \Psi _{c_{i}}\right\rangle $\ Alice
sends when he applies the bypass mode, Alice can lie about the value of the
corresponding $c_{i}$ freely, thus alters her committed $b$ in the unveil
phase.

Nevertheless, it is easy to defeat this counterfactual attack. As pointed
out in Ref. \cite{qi1025}, Bob's randomizing the optical length of path $b$
is sufficient to destroy the interference effect in the FBS system.
Therefore in our protocol, Bob can simply add extra phase shifters (other
than the one shown in FIG. 2) to both channels $X$ and $Y$ when he applies
the bypass mode, to introduce the same phase shift $e^{i\theta }$ in both
channels, where the value of $\theta $\ is randomly chosen and kept secret
from Alice, and can vary for different $\left\vert \Psi
_{c_{i}}\right\rangle $. In this case, after passing Bob's apparatus,
Alice's initial states $\left\vert \Psi _{0}\right\rangle $\ and $\left\vert
\Psi _{1}\right\rangle $\ will become, respectively,

\begin{eqnarray}
\left\vert \Psi _{0}\right\rangle  &\rightarrow &\left\vert \Psi
_{0}^{\prime }\right\rangle \equiv \sqrt{T}(e^{i\theta }\left\vert
0\right\rangle _{X})(e^{i\theta }\left\vert 1\right\rangle _{Y})  \nonumber
\\
&&-i\sqrt{R}(e^{i\theta }\left\vert 1\right\rangle _{X})(e^{i\theta
}\left\vert 0\right\rangle _{Y}),  \nonumber \\
\left\vert \Psi _{1}\right\rangle  &\rightarrow &\left\vert \Psi
_{1}^{\prime }\right\rangle \equiv \sqrt{T}(e^{i\theta }\left\vert
1\right\rangle _{X})(e^{i\theta }\left\vert 0\right\rangle _{Y})  \nonumber
\\
&&-i\sqrt{R}(e^{i\theta }\left\vert 0\right\rangle _{X})(e^{i\theta
}\left\vert 1\right\rangle _{Y}).
\end{eqnarray}

We can see that $\left\vert \Psi _{0}^{\prime }\right\rangle $\ ($\left\vert
\Psi _{1}^{\prime }\right\rangle $) differs from $\left\vert \Psi
_{0}\right\rangle $\ ($\left\vert \Psi _{1}\right\rangle $) merely by a
global factor $e^{i2\theta }$. It is well known that such a global factor is
not detectable. In fact, in our case the interference pattern of the two
wave packets meeting at the beam splitter of Alice's measurement device is
determined by their relative phase difference. No change will be detected
when they both receive a phase shift $e^{i\theta }$. Thus an honest Alice
who uses the original apparatus described in FIG. 2 will still detect $%
\left\vert \Psi _{0}^{\prime }\right\rangle $\ ($\left\vert \Psi
_{1}^{\prime }\right\rangle $) as $\left\vert \Psi _{0}\right\rangle $\ ($%
\left\vert \Psi _{1}\right\rangle $), even though she does not know the
value of $\theta $. On the other hand, if Alice wants to apply the above
counterfactual attack, without knowing $\theta $ she cannot know how to
adjust path $a$ to ensure $D_{c}$ in FIG. 3 clicking with certainty.
Consequently, there will be times that Alice does not know which mode Bob is
running. Then the number of $c_{i}$'s that she can alter will be limited,
which is insufficient to change the committed $b$\ as long as the value of $%
d/n$ in our QBC protocol is properly chosen.

\end{document}